\documentclass[%
reprint,
 amsmath,amssymb,
 aps,
pre,
floatfix,
]{revtex4-2}

\usepackage{graphicx}% Include figure files
\usepackage{dcolumn}% Align table columns on decimal point
\usepackage{subfigure}
\usepackage{bm}% bold math
\usepackage{braket}             % dirac brackets
\usepackage{amsmath,amsfonts,amssymb}
\usepackage{xcolor}
\usepackage[english]{babel}
\DeclareMathOperator{\Tr}{Tr}
\usepackage{graphicx}% Include figure files
\usepackage{dcolumn}% Align table columns on decimal point
\usepackage{bm}% bold math
\begin{document}

\preprint{APS/123-QED}

\title{Quantum operator entropies under unitary evolution}  % Force line breaks with \\

\author{Craig S. Lent}
\affiliation{%
Department of Physics\\
and Department of Electrical Engineering\\
University of Notre Dame\\
Notre Dame, IN 46556, USA
}%
%\homepage{http://www.nd.edu/~lent}

\date{\today}% It is always \today, today,

\begin{abstract}
For a  quantum state undergoing unitary Schr\"odinger time evolution, the von Neumann entropy is constant.  Yet the second law of thermodynamics, and our experience, show that entropy increases with time. Ingarden introduced the quantum operator entropy, which is the Shannon entropy of the probability distribution for the eigenvalues of a Hermitian operator. These entropies characterize the missing information about a particular observable inherent in the quantum state itself. The von Neumann entropy is the quantum operator entropy for the case when the operator is the density matrix.  We examine pure state unitary evolution in a simple model system comprised of a set of highly-interconnected topologically disordered states and a time-independent Hamiltonian. An initially confined state is subject to free expansion into available states. The time development is completely reversible with no loss of quantum information and no course graining is applied.  The positional entropy increases in time in a way that is consistent with both the classical statistical mechanical entropy and the second law.
\end{abstract}

\maketitle

% Keywords
%\keyword{entropy; quantum entropy; Landauer principle; quantum statistical mechanics}

% The fields PACS, MSC, and JEL may be left empty or commented out if not applicable
%\PACS{J0101}
%\MSC{}
%\JEL{}

%%%%%%%%%%%%%%%%%%%%%%%%%%%%%%%%%%%%%%%%%%

%%      Switch to second version to remove red highlighting
%        of changes to the manuscript
\newcommand{\MyHighlight}[1]{{{\color{red} #1}}}
% \newcommand{\MyHighlight}[1]{{{#1}}}

%\setcounter{secnumdepth}{4}
%%%%%%%%%%%%%%%%%%%%%%%%%%%%%%%%%%%%%%%%%%

\newcommand{\Lagr}{\mathcal{L}}

%%%%%%%%%%%%%%%%%%%%%%%%%%%%%%%%%%%%%%%%%%%%%%%%%%%%%%%%%%%%%%%%%%%%%%%%%%
\section{Introduction}

Entropy as a concept was defined in historical order by Clausius, Boltzmann, Gibbs, von Neumann, and Shannon. Conceptually, it would have been clearer if the order were reversed. The Shannon information-theoretic definition is the most fundamental, which can then be applied to physical quantum systems, with classical statistical mechanics following as the classical limit of the quantum case. 

Shannon chose to use the word ``entropy''  from the field of statistical mechanics for a quantity he
variously described as measuring ``choice,''  ``information'', ``uncertainty'', or ``surprise'' \cite{Shannon1948}. As the mathematical theory of communication he invented became the field of information theory, and in due course was turned back onto analyzing statistical mechanics, the layering of these various concepts often became confusing.  Information and uncertainty, for example, seem to be opposite one another. The more information one has, the less uncertainty. Whose choice is involved in the entropy of a physical system?

Ben-Naim has done the world a great favor by relentlessly clarifying the quantity defined by Shannon as a measure of the {\em missing information} associated with a probability distribution \cite{BenNaimFarewell2008}.  If all one knows is the probability distribution for a finite set of discrete possible events, the Shannon measure quantifies the amount of information, measured in bits, that one is missing. It is the difference between the incomplete knowledge captured in a probability distribution, and certainty about which event will occur. What Ben-Naim prefers to call the Shannon Measure of Information (SMI)  (or Shannon Missing Information), represents, as it were, the part of the graduated cylinder that is empty of fluid. The SMI is the more general concept; the thermodynamic entropy $S$ is a special case of the SMI applied to a particular class of physical problems.   Ben-Naim also rightly inveighs against interpreting physical entropy as a measure of ``disorder,'' a concept too vague to be scientifically quantifiable. What counts as order is entirely subjective. Ben-Naim builds on the work of Jaynes who,  reversing the historical sequence, showed specifically how statistical mechanical entropy was a particular application of Shannon's information theoretic entropy \cite{Jaynes1957a, Jaynes1957b, LentThermo2018}. 

Here we focus on the  quantum mechanical operator entropy $S_Q$,   associated with a Hermitian operator $\hat{Q}$, as formulated by Ingarden \cite{Ingarden1976} and described in Section II.  This operator entropy quantifies the amount of information about the property $Q$ that is missing in the (pure or mixed) quantum state. For example, the position operator $\hat{X}$ generates an associated entropy $S_x$ which captures how much information about position is missing. The familiar von Neumann entropy is then seen to be the special case of quantum operator entropy when the operator is the density matrix $\hat{\rho}$. Though the von Neumann entropy is zero for a pure state and constant under unitary time evolution, other operator entropies need not be. 

We examine the interpretation of these quantities in a model system with topological disorder. Section III examines the free expansion of an initially localized system, tracking several operator entropies. We see the increase in the position entropy in time which parallels classical second law behavior even in purely unitary time evolution with no coarse-graining \cite{Aguirre2019} or loss of quantum information. The position entropy saturates at levels than can be predicted by sampling random superpositions of energy eigenstates. In this way the behavior is connected to the notions of typicality that are proving so helpful in quantum statistical mechanics \cite{Gemmer2009}. In Section IV we discuss the second law of thermodynamics and time-reversibility for this system. Results for thermal equilibrium in the model system are briefly discussed in Section V. 

%%%%%%%%%%%%%%%%%%%% defining quantum operator entropy  %%%%%%%%%%%%%%%%%%%%%%%%%%
\section{Quantum operator entropy \label{sec:QuantumOperatorEntropy}}
\subsection{Definition}
We follow Ben-Naim  in defining the SMI (Shannon Measure of Information), in bits, as Shannon's measure on a probability distribution $P=[P_1, P_2, \dots, P_N]$ over $N$ possible outcomes.
\begin{equation}
    \mbox{SMI}(P)=-\sum_k^N P_k \log_2(P_k).
    \label{eq:SMIdefinition}
\end{equation}
\noindent The notation is useful here to distinguish this quantity from the physical or thermodynamic entropy $S$; it is simply a measure on a probability distribution.

To apply the Shannon measure of information in the quantum mechanical case we consider a Hermitian operator  $\hat{Q}$ written in its eigen-basis.
\begin{equation}
    \hat{Q}=\sum_k \Ket{\varphi_{q_k}}q_k \Bra{\varphi_{q_k}} 
    \label{eq:OperatorDef}
\end{equation}
\noindent We assume that the set of states  $\{ \Ket{\varphi_{q_k}} \}$ are chosen to form an orthonormal basis. 
If a measurement of $Q$ is made, the result will be one of the eigenvalues of 
$\{q_1, q_2, \dots q_k \dots q_N\}$ with probabilities $\{p_{q_1}, p_{q_2}, \dots p_{q_k} \dots p_{q_N}\}$. 
It is natural then to define the Shannon measure on this set of  probabilities as the entropy associated with $Q$.
\begin{equation}
    S_{Q}\equiv \mbox{SMI}(\left\{ p_{q_k} \right\} ) = -\sum_k p_{q_k} \log_2(p_{q_k})
     \label{eq:SQdef}
\end{equation}
\noindent This quantity was introduced by Ingarden \cite{Ingarden1976} and has been studied by Anza and Vedral, \cite{Vedral2017}, Hu {\em et al.} \cite{Hu2019}, and others.

If the system is in a pure quantum state $\Ket{\psi}$, then the probability that a measurement of $Q$ yields $q_k$ is given by the Born Rule.
\begin{equation}
    p_{q_k}=\left| \Braket{\varphi_{q_k}|\psi} \right|^2
    \label{eq:BornRule}
\end{equation}

\noindent So that
\begin{equation}
    S_{Q}(\psi)= 
     -\sum_k \left| \Braket{\varphi_{q_k}|\psi} \right|^2
     \log_2\left(\left| \Braket{\varphi_{q_k}|\psi} \right|^2) \right).
     \label{eq:SQpure}
\end{equation}

Entropy quantifies missing information, so what information is missing? For an observable $Q$ and a pure state $\Ket{\psi}$, $S_Q$ measures the number of bits of information that are missing {\em from the universe} concerning what value of $Q$ will be obtained if a measurement of $Q$ is made. The quantum state of the system $\Ket{\psi(t)}$ contains everything there is to know about the system at time $t$. Because of fundamental quantum indeterminism, that is not enough to pin down which eigenvalue of $\hat{Q}$ will be measured (unless, of course, $\Ket{\psi}$ happens to be an eigensate of $\hat{Q}$). We now know from recent Bell test experiments that this indeterminism is a fundamental feature of  reality \cite{Shalm2015, Giustina2015, Hensen2015, Zeilinger2018}. It is not just a feature of quantum mechanics as we currently understand it, nor is it just an expression of the limited information of an observer. The quantum operator entropy $S_Q$ therefore reflects information about the observable Q that is missing from the physical world. It is, in that somewhat strange sense, an {\em objective} property of the physical system and not a {\em subjective} property of the observer's knowledge. Perhaps one can speak of Shannon's notion of ``choice'' applying here. When confronted with a measurement of $Q$, the physical world has $S_Q$ bits of choice in the outcome that are unconstrained by the physical law, a law which has determined the present $\Ket{\psi(t)}$. The missing information about $Q$ in (\ref{eq:SQpure}) is really missing. 

A mixed state describes a system $S$ which  is either coupled dynamically to a reservoir system $R$ or has been so in the past. The quantum state of the composite system is not in general simply a direct product of the state of each subsystem but rather an entangled state $\Ket{\psi^{SR}}$. The density operator for the composite system is defined by
\begin{equation}
    \hat{\rho}^{SR}= \Ket{\psi^{SR}}\Bra{\psi^{SR}}.
    \label{eq:rhoOuterProduct}
\end{equation}
\noindent The density operator for the system alone is defined using the partial trace over the reservoir degrees of freedom.
\begin{equation}
     \hat{\rho}\equiv  \Tr_R \left( \hat{\rho}^{SR}\right) 
     \label{eq:RhoReduced}
\end{equation}

We can write the density operator in the basis of its own eigenstates. 
\begin{equation}
    \hat{\rho}=\sum_k \Ket{\nu_k} \rho_k \Bra{\nu_k}.
    \label{eq:RhoEigenbasis}
\end{equation}
\noindent For a pure state only one of the $\rho_k$'s is nonzero.

The probability $p_{q_k}$ that a measurement of $Q$ for the system yields $q_k$ can  be calculated for the mixed state using the density operator and the projection operator onto the $k^{th}$ eigenstate of $\hat{Q}$.
\begin{equation}
     p_{q_k}= \Tr\left(\hat{\rho} \; \left[\;\Ket{\phi_{q_k}}\Bra{\phi_{q_k}}\;\right] \right)
     \label{eq:ProbFromRho}
\end{equation}
\noindent The quantum operator entropy for a mixed (or pure) state is then given by applying (\ref{eq:SQdef}) to (\ref{eq:ProbFromRho}).
\begin{multline}
     S_Q(\hat{\rho}) \equiv  \\ 
     -\sum_{k}\Tr\left( \hat{\rho} \; \left[\;\Ket{\phi_{q_k}}\Bra{\phi_{q_k}}\;\right] \right)
     \log_2\left( \Tr\left( \hat{\rho} \; \left[\;\Ket{\phi_{q_k}}\Bra{\phi_{q_k}}\;\right] \right)\right)
     \label{eq:SQFromRho}
\end{multline}
\noindent The expression in (\ref{eq:SQFromRho}) includes the previous expression in (\ref{eq:SQpure}) as a special case when $\rho$ represents a pure state. 

Again, $S_Q$ is providing a measure (in bits) of missing information. For a  mixed state, the source of this missing information is two-fold. Quantum indeterminacy still limits the information about a future measurement that is present in the current state of the system. But in addition there is also information missing about the reservoir's state and the mutual information characterizing the entanglement between system and reservoir. The system's reduced density matrix $\hat{\rho}$ is not a complete description of the quantum state of the system, but it is the best possible local description.  For the physical world, there is a fact-of-the-matter about the global quantum state that includes both system and reservoir $\Ket{\psi^{SR}}$. But the local description of the system alone represented by $\hat{\rho}$ has less information. 

%%%%%%%%%%%%%%%%%%%% examples of quantum operator entropy  %%%%%%%%%%%%%%%%%%%%%%%%%%
\subsection{Examples of quantum operator entropies}

If we use a basis set of discrete position eigenstates $\Ket{x_k}$ we can define the quantum operator entropy for $\hat{X}$, the position operator. 
\begin{equation}
     S_x(\hat{\rho}) = -\sum_{k}\Tr\left( \hat{\rho} \; \left[\;\Ket{{x_k}}\Bra{{x_k}}\;\right] \right)
     \log_2\left( \Tr\left( \hat{\rho} \; \left[\;\Ket{{x_k}}\Bra{x_k}\;\right] \right)\right)
     \label{eq:SX}
\end{equation}

\noindent The quantum operator entropy for the Hamiltonian $\hat{H}$ with allowed energies $E_k$ and eigenstates $\Ket{E_k}$ is

\begin{equation}
     S_E(\hat{\rho}) = -\sum_{k}\Tr\left( \hat{\rho} \; \left[\;\Ket{{E_k}}\Bra{{E_k}}\;\right] \right)
     \log_2\left( \Tr\left( \hat{\rho} \; \left[\;\Ket{{E_k}}\Bra{E_k\;}\right] \right)\right)
     \label{eq:SH}
\end{equation}

\noindent The quantum operator entropy for the density operator $\hat{\rho}$ itself, from (\ref{eq:RhoEigenbasis}) is 
\begin{equation}
     S_\rho(\hat{\rho}) = -\sum_{k}\Tr\left( \hat{\rho} \; \left[\;\Ket{{\nu_k}}\Bra{{\nu_k}}\;\right] \right)
     \log_2\left( \Tr\left( \hat{\rho} \; \left[\;\Ket{{\nu_k}}\Bra{\nu_k}\;\right] \right)\right)
     \label{eq:Srho}
\end{equation}
\noindent Because 
\begin{equation}
      \Tr\left( \hat{\rho} \; \left[\;\Ket{{\nu_k}}\Bra{{\nu_k}}\;\right] \right)= \rho_k,
     \label{eq:rhoProjDiag}
\end{equation}
 we can write 
\begin{equation}
     S_\rho(\hat{\rho}) = -\sum_{k} \rho_k \log_2(\rho_k)= \mbox{SMI}(\{\rho_k\}).
     \label{eq:SrhoShannon}
\end{equation}
\noindent The quantum operator entropy for the density operator is the SMI of the diagonal elements of the density matrix. 
We recognize   (\ref{eq:SrhoShannon}) as the expectation value of $-\log_2(\hat{\rho})$, and so write
\begin{equation}
    S_\rho(\hat{\rho})
    =\Braket{-\log_2(\hat{\rho}) } 
    =- \Tr\left( \hat{\rho} \log_2(\hat{\rho}) \right) 
    =S_{vN}(\rho).
         \label{eq:Svn}
\end{equation} 
\noindent The entropy $S_\rho$ is identical to the von Neumann entropy $S_{vN}$ (in bits). For a pure state,  the entropy $S_\rho=S_{vN}$ is zero.

In general each operator entropy $S_Q$ can take on different values because each quantifies something different. If what is known about the system is $\hat{\rho}$,  $S_x(\hat{\rho})$ is the amount of information that is missing about position, or more precisely, about the outcome of position measurements. It is the answer to the question: How many bits of information are not known about the outcome of a position measurement if all one knows is $\hat{\rho}$?  $S_E(\hat{\rho})$ is the amount of information about energy that is missing.  $S_\rho(\hat{\rho})$ is the amount of information that is missing about which quantum state the system will be found in. It measures the ``mixedness'' or purity of the state.  $S_\rho(\hat{\rho})$, the von Neumann entropy, is invariant under unitary transformations of the basis. Other operator entropies $S_Q(\hat{\rho})$ are not invariant---they are tied to the eigen-basis of the particular operator $\hat{Q}$.

%%%%%%%%% Entropy change of pure state under unitary evolution  %%%%%%%%%%%%%%%
\section{Entropy change of a pure state under unitary evolution}
\subsection{Model system}
We consider here a model problem of unitary evolution of a pure state in a closed system with a time-independent Hamiltonian.  The purpose is to examine the different roles played by the von Neumann entropy $S_{vN}=S_\rho$, the energy entropy $S_E$, and the positional entropy $S_x$. 

\begin{figure}[t]
\centering
\includegraphics[width=8.6cm]{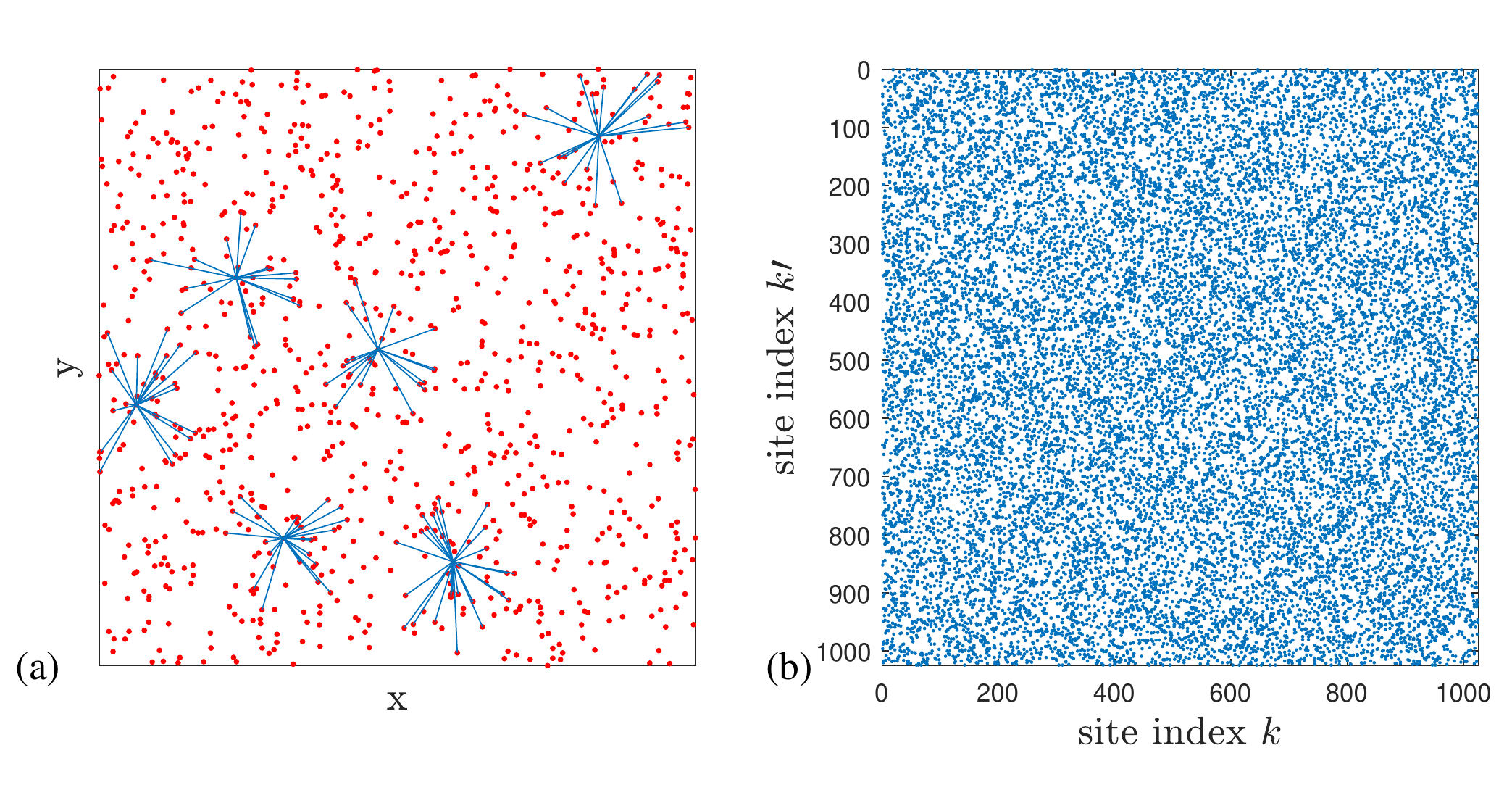}
\caption{ The geometry and connectivity of a model system with 1024 randomly positioned sites.
(\textbf{a}) The position of each site $k$ is indicated as a  dot at $\mathbf{r}_k$.  For six representative sites, the  connections to nearby sites are indicated by lines. 
For each site, connections are made to a randomly chosen subset of the 50 nearest other sites. 
(\textbf{b}) Site-to-site connectivity is shown by a  dot on the $k_{th}$ row and $k'_{th}$ column if the  site $k$ is connected to site $k'$. A connection means that there is a corresponding nonzero Hamiltonian matrix element
$\Braket{\mathbf{r}_k|\hat{H}|\mathbf{r}_{k'}}=\gamma_0$.  The number of connections for each site (the number of  dots in a row or column) varies between a minimum of 10 and a maximum of 32, with a mean value of 18.
}
\label{fig:RandomSiteGeometry}
\end{figure}   

The system consists of a random array of $N=1024$ fixed sites. Each site is labeled with an index $k =[1,2, 3, \dots, N]$ and is at a randomly chosen position $\mathbf{r}_k=(x_k,y_k)$ a unit square. The site positions are shown graphically in Figure \ref{fig:RandomSiteGeometry}a as  dots. We use as basis states for the system the set of all the position eigenstates localized on each site, $\{\Ket{\mathbf{r}_k}\}$. The on-site energy for each is $E_0$.

Off-diagonal elements of the Hamiltonian couple each site to several nearby sites with a fixed coupling matrix element $\gamma_0 = \Braket{\mathbf{r}_k|\hat{H}|\mathbf{r}_{k'}}$. Each site is coupled to a randomly chosen subset of its 50 nearest sites. The  coupling to neighbors for six representative sites is shown in Figure \ref{fig:RandomSiteGeometry}a as  lines connecting the dots.  Figure \ref{fig:RandomSiteGeometry}b shows the connectivity of the sites with a  dot on row $k$, column $k'$, if site $k$ is coupled to site $k'$.

The Hamiltonian for the system is 
\begin{equation}
    \hat{H}= \sum_k \Ket{\mathbf{r}_k}E_0\Bra{\mathbf{r}_k}
                -  \sum_{k,k'} \gamma(\mathbf{r}_k,\mathbf{r}_{k'})\left [\Ket{\mathbf{r}_k}\Bra{\mathbf{r}_{k'}}
                 + \Ket{\mathbf{r}_{k'}}\Bra{\mathbf{r}_k}\right]
    \label{eq:RandomNetworkH}
\end{equation}
\noindent Here $\gamma(\mathbf{r}_k,\mathbf{r}_{k'})=\gamma_0$ if two sites are connected, and 0 if they are not. 
% The eigenvalues of $\hat{H}$ are denoted $E_j$.

The  randomness in this model minimizes the artifacts of geometric regularity on the dynamics. We want to see how these different operator entropies change due to fundamental unitary dynamics without the patterns of constructive and destructive interference that dominate, for example, the evolution of a similar system on a regular lattice. This topologically disordered model is similar to, but distinct from, the Lifshitz model for disordered semiconductors \cite{Lifshitz1964} for which $\gamma$ is simply a function of the distance between sites. The high multiple connectivity also lets us generalize the interpretation of the model as will be discussed below.

To construct the connections between sites, nine passes through all the sites are made, adding a connection between each site and another site randomly chosen from among its 50 closest neighbors. The result is that the  number of connections for each site varies between 10 and 32, with a mean of 18. (The situation is complicated by the fact that site $k$ might have site $j$ as one of its 50 nearest neighbors, while site $j$ does not have site $k$ as one of its 50 nearest neighbors.)

\subsection{Unitary free expansion}
\begin{figure}[tb]
\centering
\includegraphics[width=8.6cm]{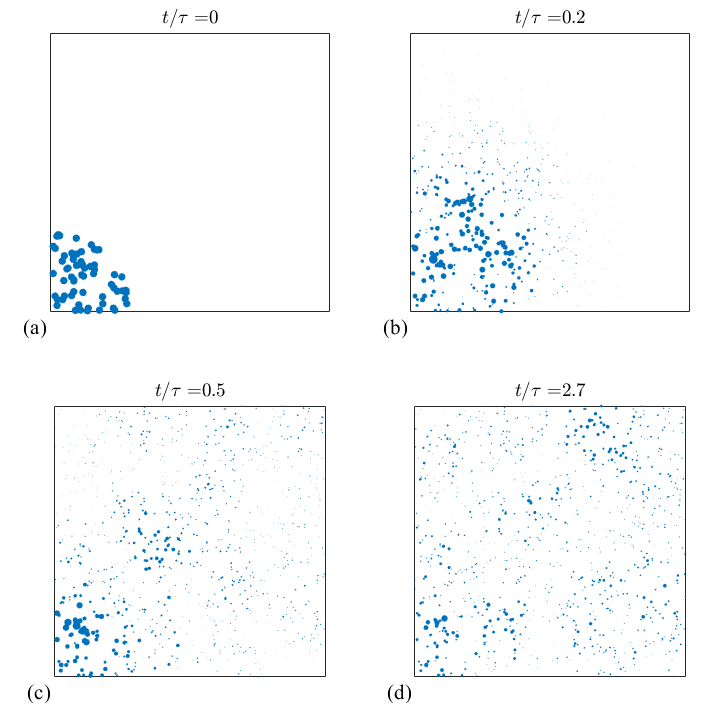}
\caption{Expansion of the probability distribution under unitary evolution for the system shown in Figure \ref{fig:RandomSiteGeometry}. The initial state is a pure quantum state with uniform probability over the 64 sites closest to the origin (lower left of each panel). The time development is calculated from equation (\ref{eq:UnitaryTimeEvolutionPsi}). 
The system is at all times isolated and remains in a pure quantum state. Panels (\textbf{a}-\textbf{d}) show snapshots of the probability distribution at various times.  The time scale is measured in units of $\tau=\pi \hbar/ \gamma_0$. The area of each dot is proportional to the probability of the system being found on that site. The system as modeled cannot dissipate energy so quantum interference effects persist and the distribution will  never become completely homogeneous. The evolution is reminiscent of the free expansion of an ideal gas, but because it is unitary, the von Neumann entropy is constant and the motion is entirely reversible. 
}
\label{fig:ExpansionSnapshots}
\end{figure}

The dynamic problem we solve is the expansion of the state from an initially  spatially confined state. The initial state has an equal probability distributed among the 64 sites that are closest to the origin. We solve for the time development of the state function using the unitary time development operator.

\begin{equation}
    \Ket{\psi(t)} = e^{-i \hat{H} t/\hbar} \Ket{\psi(0)}
    \label{eq:UnitaryTimeEvolutionPsi}
\end{equation}
\noindent The expansion of the state into the surrounding state space is shown by the snapshots of the probability density in Figure \ref{fig:ExpansionSnapshots}. The time scale is set by the characteristic tunneling time between connected sites 
\begin{equation}
    \tau=\pi \hbar/ \gamma_0. \label{eq:TauDef}
\end{equation}
 The figure shows how the probability expands much like a classical gas into the available states. Since the expectation value of the energy is constant during unitary evolution, the system cannot de-excite, and quantum interference fluctuations persist indefinitely. 

%%%%%%%%%%% Figure 2  Expansion snapshots  %%%%%%%%%%%%%%%
%%%%   %%%%%%%%%%%%%%%  %%%%%%%%%%%%%%%  %%%%%%%%%%%%%%%

\begin{figure}[bt]
\centering
\includegraphics[width=8.6cm]{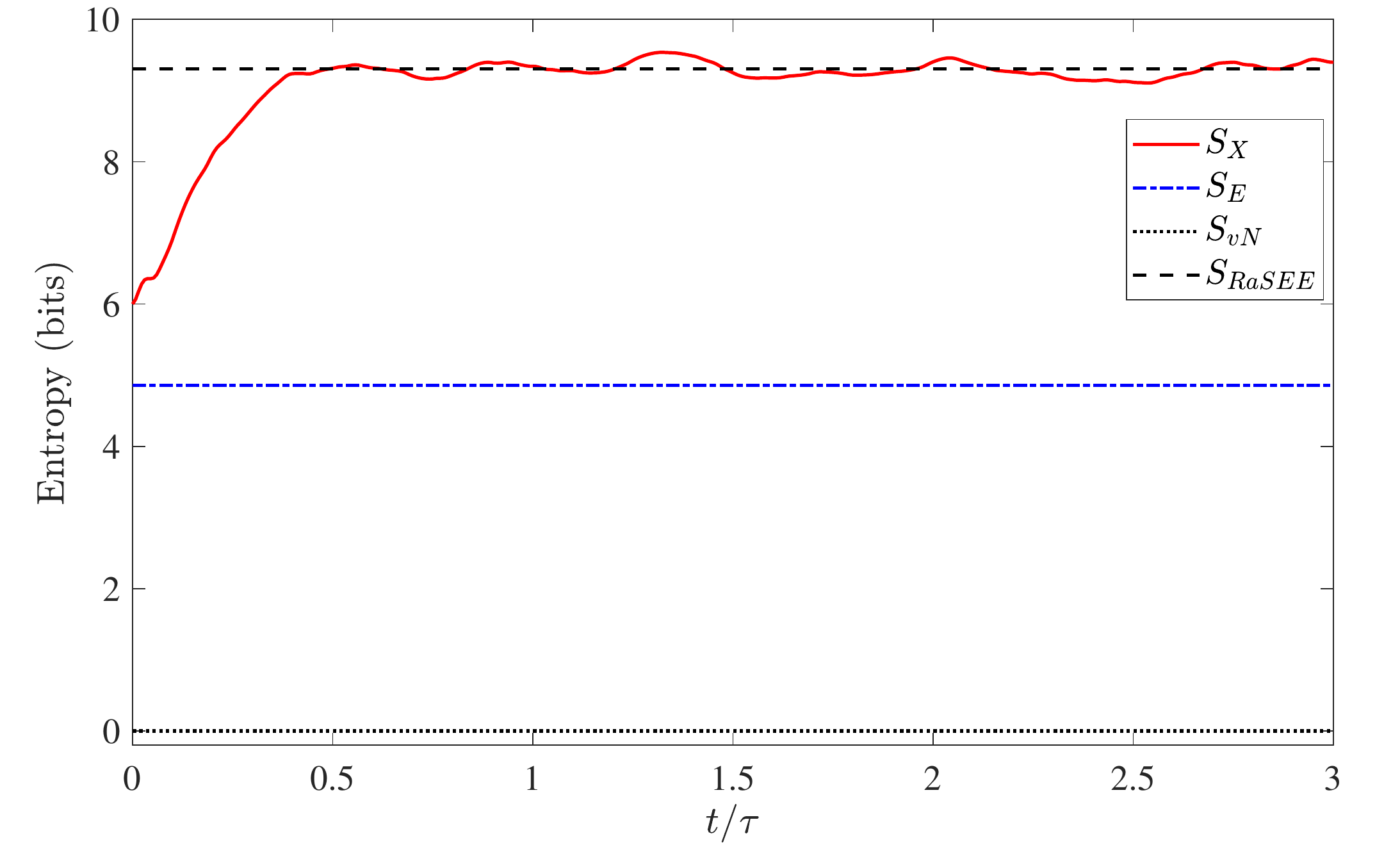}
\caption{Quantum operator entropies for the unitary expansion  shown in Figure \ref{fig:ExpansionSnapshots} on the geometry shown in Figure \ref{fig:RandomSiteGeometry}. The system evolves from the localized state of  Figure \ref{fig:ExpansionSnapshots}a according to equation (\ref{eq:UnitaryTimeEvolutionPsi}) with the Hamiltonian given by (\ref{eq:RandomNetworkH}).  The time scale is measured in units of $\tau=\pi \hbar/ \gamma_0$. The von Neumann entropy (dotted line) is zero throughout because it is a measure of the purity or ``mixed-ness'' of the quantum state. The state here is always a pure quantum state and the evolution is reversible. The energy entropy $S_E$ is a measure of the amount of missing information about the state's energy. Because the state is not a stationary state, but rather a linear combination of stationary states, several outcomes of an energy measurement are possible. The value of $S_E$ is the answer to the question: How many bits of information about the results of an energy measurement are missing, if all one knows is the quantum state $\Ket{\psi(t)}$? Here the answer is 4.86 bits, and is constant in time because unitary time evolution preserves the values of projections onto energy eigenstates. The value of the position entropy $S_x$ (solid line) is similarly the amount of missing information about the outcomes of a position measurement, given the quantum state. The position entropy  resembles the behavior of the classical entropy of an ideal gas which increases logarithmically  with volume.   
The dashed line shows the average value of $S_x$ over 300 samples of random superpositions of energy eigenstates (RaSEE), as described in Section \ref{sec:RaSEE}, with a value of 9.39 bits. 
}
\label{fig:ExpansionEntropies}
\end{figure}

Figure \ref{fig:ExpansionEntropies} shows the the calculated entropies $S_x$, $S_E$, and $S_{vN}=S_\rho$, the von Neumann entropy, during the expansion shown in Figure \ref{fig:ExpansionSnapshots}. The von Neumann entropy is, of course, constant during the unitary evolution and is in fact 0 because the state is always pure. 

%%%%%%%%%%% Figure 3  Entropies during expansion (64 init)  

\begin{figure}[hbt]
\centering
\includegraphics[width=8.6cm]{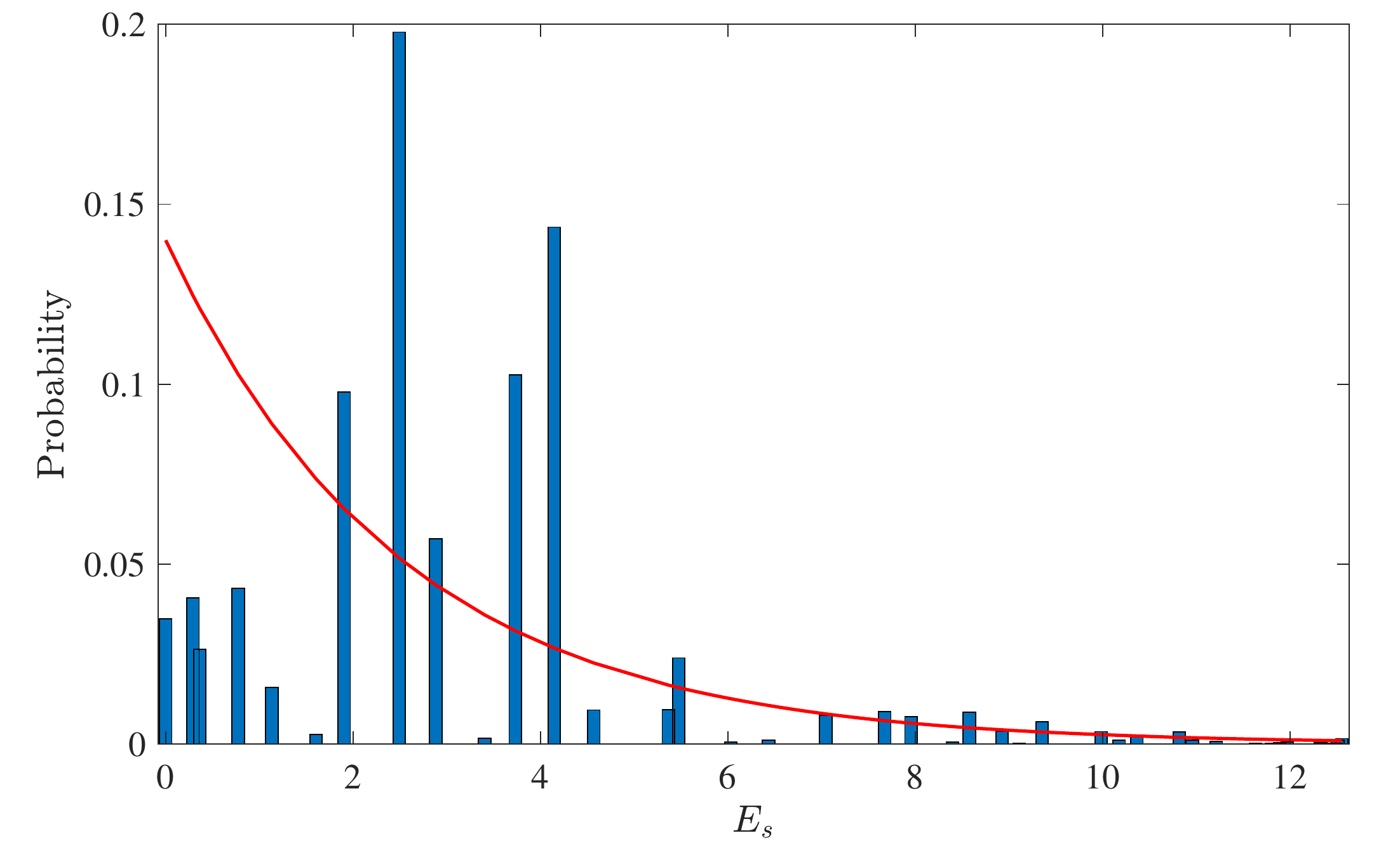}
\caption{Probability distribution in energy. For the  state shown in Figure \ref{fig:ExpansionSnapshots}, the probability of finding the system in the $k^{\mbox{}th}$ energy eigenstate, with scaled energy eigenvalue $E^k_s=(E_k-E_1)/\gamma$,  is shown by by the bar chart for the lowest 50 eigenstates. These probabilities do not change in time during  unitary evolution. The position entropy $S_x$ shown in  Figure \ref{fig:ExpansionEntropies} rises in the expansion to a roughly steady-state level, and this might be described as ``thermalization,'' but the probability distribution in energy is clearly not that of a thermal equilibrium state. The line is a thermal Boltzmann distribution shown for comparison. The energy entropy $S_E$ is the Shannon entropy of this probability distribution. 
}
\label{fig:Penergy64}
\end{figure}

The energy entropy $S_E$ is also constant during the expansion, but it is not $0$. The energy eigenstate occupation probabilities cannot change during unitary time development so $S_E$ is independent of time. $S_E$ is  4.86 bits (rather than 0) because the initial state is not a Hamiltonian eigenstate, so there are many energy eigenvalues that could be measured. $S_E$ characterizes the missing information in the probability distribution of those energy measurements, shown in Figure \ref{fig:Penergy64}. 

The positional entropy $S_x$ characterizes the missing information about position. At $t=0$, it is exactly 6 bits because the probability is uniformly distributed among $64=2^6$ sites. As the expansion progresses it increase to a value between $9$ and $10$. If the distribution were distributed completely evenly  among the $1024=2^{10}$ sites, $S_x$ would be 10. The initial quantum confinement means that the isolated system is excited and it has no way of de-exciting. If it did, $S_x$ would approach a value of 10 bits. Nevertheless, it is clear that the increase in the quantum mechanical measure $S_x$  resembles the increase in the classical thermodynamic entropy of  an  ideal gas for which $\Delta S = \log_2(V_f/V_i)$ bits. Expanding the volume by a factor of 16 would increase the classical statistical mechanical entropy by $\Delta S = \log_2(16)=4$ bits. The dashed line in Figure \ref{fig:ExpansionEntropies} is the position entropy of a random superpostition of energy eigenstates, as discussed further in Section \ref{sec:RaSEE}.

We emphasise that despite the increase in position entropy $S_x$ shown in Figure \ref{fig:ExpansionEntropies}, the system is evolving in a completely reversible way. No information about the quantum state is being lost.
%-- removed there is no coarse graining in this calculation.  
Equation (\ref{eq:UnitaryTimeEvolutionPsi}) can be inverted so we could use the state $\Ket{\psi(t)}$ at any time $t$ to reconstruct precisely the initial state $\Ket{\psi(0)}$. The constant purity of the state is precisely reflected in the unchanging value of the von Neumann entropy $S_{vN}$. What {\em is} changing is the amount of missing information in the quantum state {\em about position}, and precisely that is quantified by $S_x$.  

The Hamiltonian for the system given by (\ref{eq:RandomNetworkH}) has eigenvalues $E_k$. It is helpful to scale energies relative to the ground state in units of $\gamma$.
\begin{equation}
     E_s = (E - E_1)/\gamma. \label{eq:Escaling}
\end{equation}
The scaled energy eigenvalues for the particular Hamiltonian (i.e., random interconnections) shown in Figures \ref{fig:RandomSiteGeometry}, \ref{fig:ExpansionSnapshots}, and \ref{fig:ExpansionEntropies} extend from $E_s^1=0$ to $E_s^{1024}=26.9$. The expectation value of the scaled energy  is $\left<E_s\right>=5.12$, and is independent of time. This energy reflects the kinetic energy of confining the system into the initial $N_{init}=64$ states. 

%
%  for this case, S_E= 4.86 bits ??
%

One might describe the apparent saturation of $S_x(t)$ as the system ``thermalization,'' but it is important to note that it is not in a thermal equilibrium state. Figure \ref{fig:Penergy64} shows the probability distribution for the 50  lowest  eigenenergies. The line is a thermal Boltzmann distribution. Obviously the state is far from being thermal in energy. Because the time evolution is unitary, this probability distribution is constant in time. 

%%%%%%%%%%% Figure 4  Probability distribution in energy  
%%%%%%%%%%%  %%%%%%%%%%% %%%%%%%%%%% %%%%%%%%%%% 

%%%%%%%%%%%%%%%%%%%%%%%%%%%%%%%%%%%%%%%%%%%%%%%%%%
%%%%%%%%%%%%%%%%%%%%%%%%%%%%%%%%%%%%%%%%%%%%%%%%%%
\subsection{Differing random configurations}

\begin{figure}[hbt]
\centering
\includegraphics[width=8.6cm]{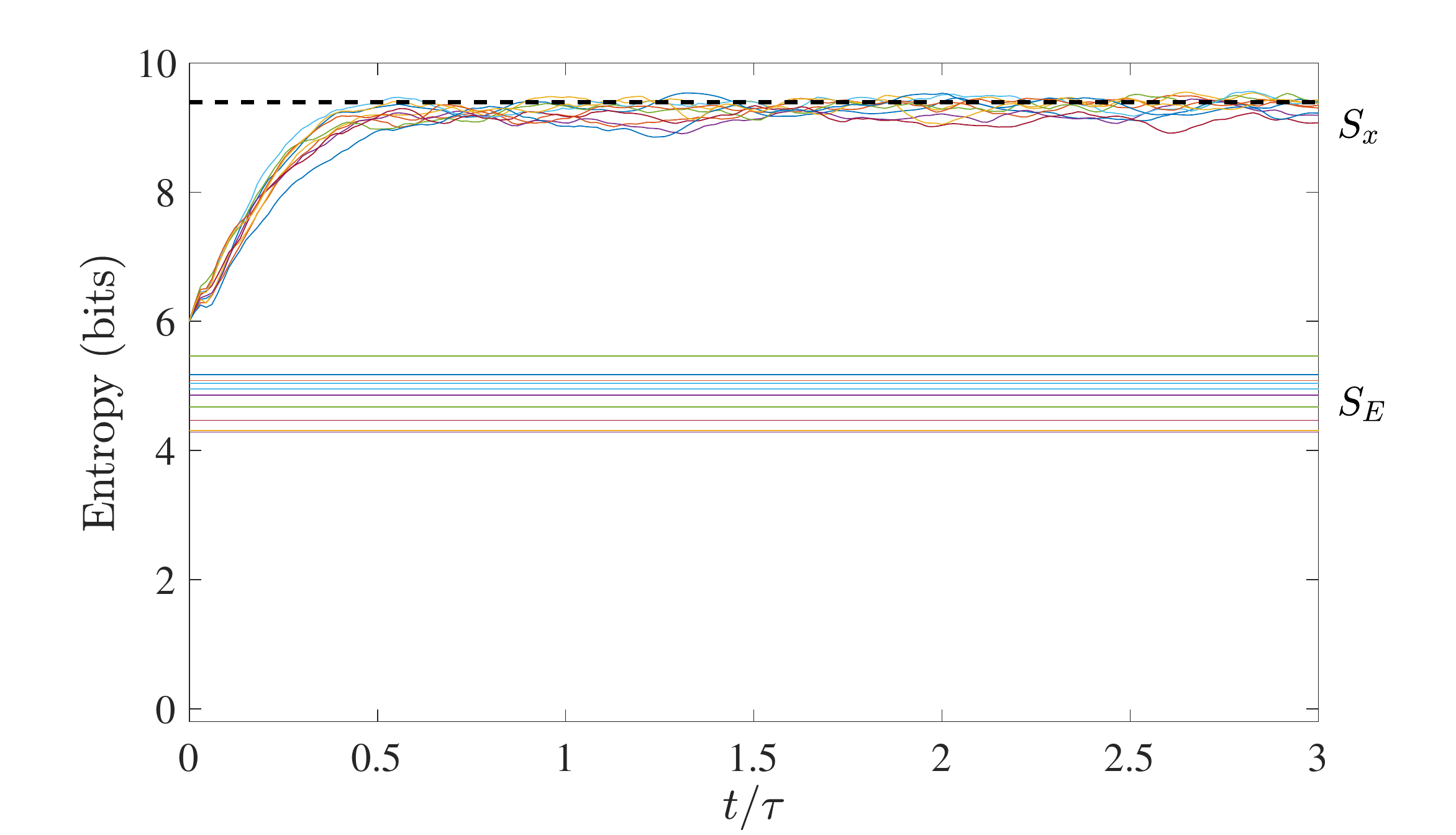}
\caption{Entropies during free expansion for different random connectivities. 
The evolution of entropies $S_x(t)$ and $S_E(t)$ are shown for free expansion from an initial state with uniform occupancy over 64 sites, as in Figure \ref{fig:ExpansionEntropies}, but with random local interconnections that differ in detail. Results from 10 different randoms configuration are shown, each similar to that illustrated in Figure \ref{fig:RandomSiteGeometry}. The Hamiltonian is given by (\ref{eq:RandomNetworkH}) and the time evolution is unitary. The total system size is $N=1024$.  The von Neuman entropy is always 0. The energy entropy $S_E$ is  constant in time but assumes different values for different connectivity configurations. The eigenvalue spectrum for each case is distinct.  The position entropy $S_x$ rises from 6 bits to more than 9 bits. Once near the saturation value, differences between different configurations are comparable in magnitude to the the quantum oscillations for each case. The dashed line shows the average value of $S_x$ over 300 samples of random superpositions of energy eigenstates (RaSEE), as described in Section \ref{sec:RaSEE}, with a value of 9.39 bits. 
}
\label{fig:SMulticonfig64}
\end{figure}

The specific configuration of the random local connectivities of the Hamiltonian (\ref{eq:RandomNetworkH}) affects the details of $S_x(t)$, but not the overall shape of the saturation to a typical value.
Figure \ref{fig:SMulticonfig64} shows $S_x(t)$ and $S_E(t)$ for 10 different random configuration, chosen in the same way as describe above, with the initial state localized uniformly across  the 64 sites nearest to the origin (hence $S_x(0)=6$). Because of the differences in (\ref{eq:RandomNetworkH}), the scaled energy expectation values $\left< E_s \right>$ vary  between 3.38 and 5.88 across this set of configurations. The energy operator entropy $S_E$ varies similarly between 4.28 and 5.46 bits and are, of course, constant in time. 
The position entropy $S_x$ in each case increases and saturates around the same values, despite the difference in the configurations. The variations in time due to quantum interference fluctuations are of the same magnitude as the differences between different configurations. The dashed line in Figure \ref{fig:SMulticonfig64} shows the position entropy for the average value of $S_x$ over 300 random superpositions of the energy eigenstates (RaSEE) for the first of the ten Hamiltonians, $S_x=9.39$ bits. The exact value so obtained varies slightly with the specifics of each configuration because the eigenvalue spectrum of each is different in detail. But again, the variation across configurations is comparable in magnitude to that of the quantum fluctuations in time. 

%%%%%%%%%%% Figure 5  Multiple configurations N=64 
%%%%%%%%%%%  %%%%%%%%%%% %%%%%%%%%%% %%%%%%%%%%% 

%%%%%%%%%%% Figure 6  Differing initial confinements  
\begin{figure}[hbt]
\centering
\includegraphics[width=8.6cm]{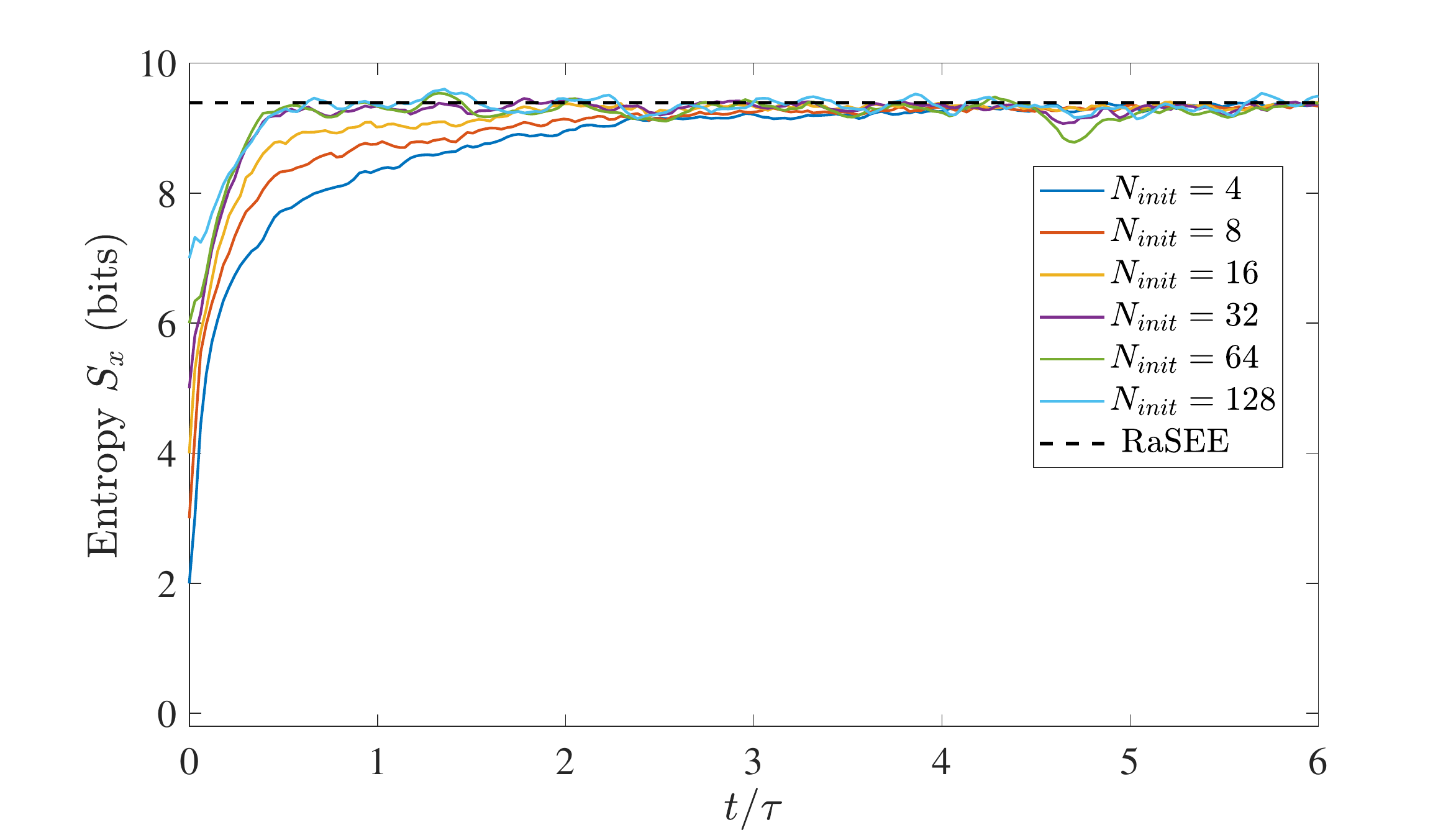}
\caption{The position entropy $S_x$ during free expansion for various initial confinements. The initial wavefunction was uniformly distributed over the $N_{init}$ sites closest to the origin. Figure \ref{fig:ExpansionSnapshots}a shows the initial state for $N_{init}=64$. Here we see the results for $N_{init} =4, 8, 16, 32, 64, 128$, corresponding to $S_x(0)= 2, 3, 4, 5, 6$, and 7 bits (different curves can be identified by the value at $t=0$). For each of these cases the connectivity configuration, and therefore the Hamiltonian, was the same as that for Figures \ref{fig:ExpansionSnapshots}, \ref{fig:ExpansionEntropies}, and \ref{fig:Penergy64}.  The dashed line shows the average value of $S_x$ over 300 samples of random superpositions of energy eigenstates (RaSEE), as described in Section \ref{sec:RaSEE}, with a value of 9.39 bits. 
}
\label{fig:SxMultiNinits}
\end{figure}   
%%%%%%%%%%%  %%%%%%%%%%% %%%%%%%%%%% %%%%%%%%%%% 

%%%%%%%%%%%%%%%%%%%%%%%%%%%%%%%%%%%%%%%%%%%%%%%%%%
%%%%%%%%%%%%%%%%%%%%%%%%%%%%%%%%%%%%%%%%%%%%%%%%%%
\subsection{Differing initial confinements}

Figure \ref{fig:SxMultiNinits} shows the position entropy $S_x(t)$ under unitary time evolution for different initial  states. The Hamiltonian (and random interconnectivity) is  identical to  that which produced the time development shown in Figures \ref{fig:ExpansionSnapshots} and \ref{fig:ExpansionEntropies}. The initial states  are in each case  chosen to be a uniform distribution of the $N_{init}$ sites nearest the origin where
\begin{equation}
N_{init} =4, 8, 16, 32, 64, 128.
\end{equation}
 The corresponding values of $S_x(0)$ are 2, 3, 4, 5, 6, and 7 bits. In contrast to the situation of Figure \ref{fig:SMulticonfig64}, the Hamiltonian here is exactly the same for all cases; only the initial condition is varied.
The tightly confined states have a higher energy expectation value. For  these initial states with $N_{init}$ from 4 to 128 we have
\begin{equation}
    \left< E_s \right> = 18.2, 16.4, 14.2, 8.9, 5.11, 4.52. 
    \label{eq:EsConfined}
\end{equation}
The behavior of $S_x(t)$ is basically the same in each case, increasing to a value of about 9.4 bits,  with persistent quantum oscillations, as the probability expands to fill the available state space.  It is notable that the systems saturate to roughly the same value, though the energy expectation values are very different.

The dashed line in Figure \ref{fig:SxMultiNinits} shows the  RaSEE position entropy, averaged over 300 random superpositions of energy eigenstates, with a resulting value of $9.39\pm 0.02$. For the ensemble of random superpositions, the expectation values of the energy are high compared with (\ref{eq:EsConfined}), averaging  $\left< E_s \right> = 18.7 \pm 0.2$. Nevertheless the RaSEE value is remarkably consistent with the saturation value of $S_x$ regardless of the initial confinement.

%%%%%%%%%%%%%%%%%%%%%%%%%%%%%%%%%%%%%%%%%%%%%%%%%%
%%%%%%%%%%%%%%%%%%%%%%%%%%%%%%%%%%%%%%%%%%%%%%%%%%
\subsection{Random superpositions of energy eigenstates (RaSEE) \label{sec:RaSEE}}

We have seen in our model system that for different initial confinements and for different random connectivities, unitary time-evolution (\ref{eq:UnitaryTimeEvolutionPsi}) drives the the position entropy $S_x(t)$ toward   the same average value, about 9.39 bits, though with small and persistent quantum oscillations. How could this value be predicted prior to solving the detailed dynamics of (\ref{eq:UnitaryTimeEvolutionPsi})?

\begin{figure}[hbt]
\centering
\includegraphics[width=8.6cm]{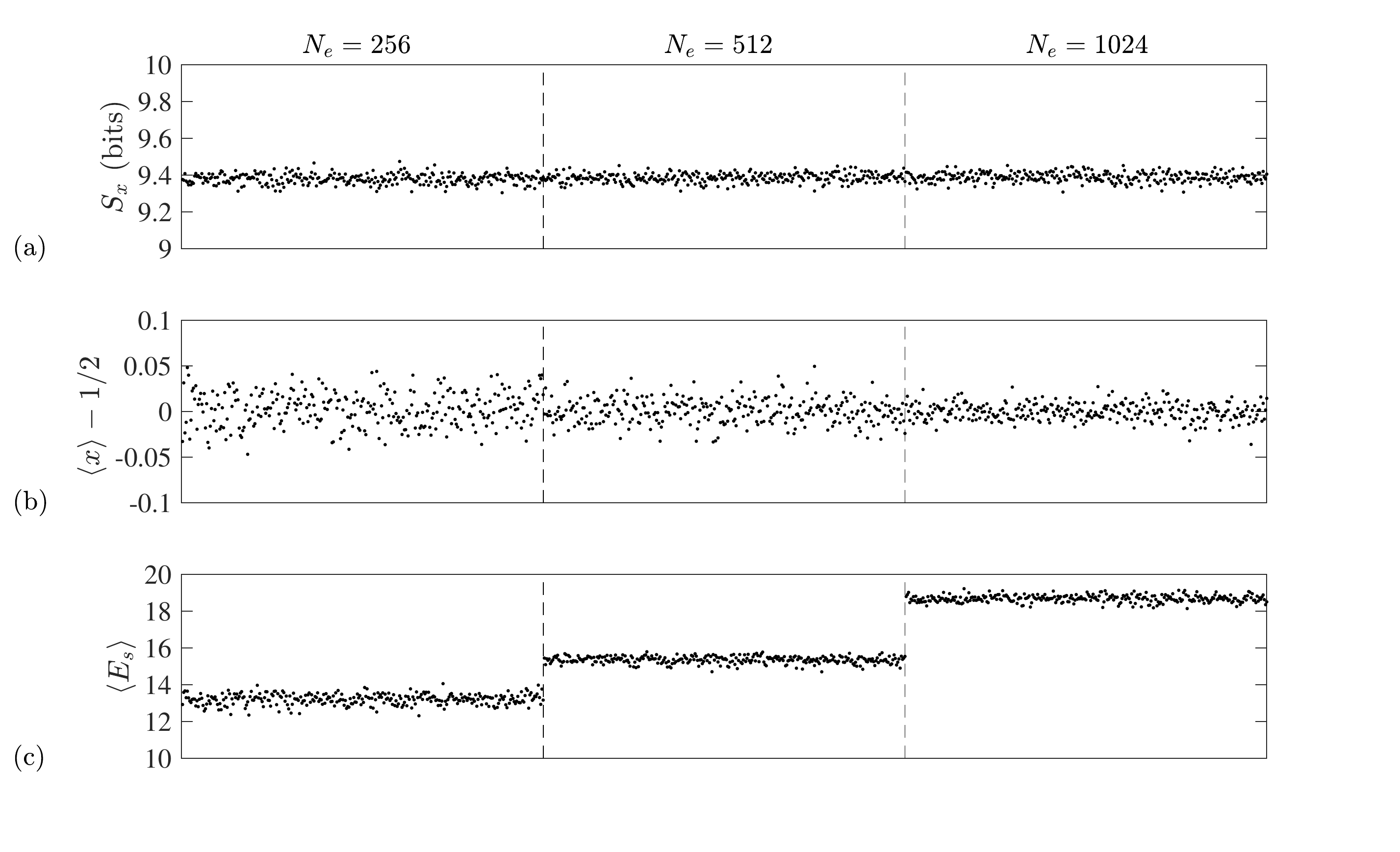}
\caption{Properties of random superpositions of energy eigenstates (RaSEE). A RaSEE state is constructed by taking a randomly weighted sum of the first $N_e \le N$ Hamiltonian eigenstates with random phases as defined by equation (\ref{eq:DefRaSEE}). 
We calculate 900 sample RaSEE states. 
(a) The values of $S_x$ for the RaSEE states. The average value of $S_x$ for RaSEE states is 9.39 $\pm$ 0.02 bits. This value matches the approximate saturation value of $S_x(t)$ obtained by unitary time evolution as shown by dashed lines in Figures \ref{fig:ExpansionEntropies}, \ref{fig:SMulticonfig64}, and \ref{fig:SxMultiNinits}.
(b) The expectation values of position in the unit interval. Here $\left< x \right>=0.5$ corresponds to the center of the square shown in Figures \ref{fig:RandomSiteGeometry}a and  \ref{fig:ExpansionSnapshots}. (c) The scaled expectation value of the energy 
$\left<E_s\right>=(\Braket{E}-E_1)/\gamma$, 
where $E_1$ is the ground state. For each quantity we show 300 samples each for the cases when $N_e$ is 256, 512, and 1024 (the full spectrum). Increaseing $Ne$ results in an upward shift in the  energy expectation values,  a smaller variance in $\Braket{x}$, and no apparent change in $S_x$.  
}
\label{fig:ExpRaSEE}
\end{figure}

For a given N-dimensional Hamiltonian $H$ with eigenvalues $E_k$ and eigenstates $\Ket{E_k}$, we can construct a random superposition of energy eigenstates (RaSEE). This involves choosing points randomly and uniformly distributed on the surface of the unit sphere in a Hilbert space of $N_e$ dimensions ($N_e \le N$). To do this we use the method of Marsaglia \cite{Marsaglia1972}.  We first construct a  vector $\mathbf{w}$ of length $N_e$  whose components $w_k$  are random normal deviates with unit variance. The vector is then normalized to unit length,
$\mathbf{w'}=\mathbf{w}/\| \mathbf{w}\|$, and
random phase factors $\phi_k$ are chosen from a uniform distribution over the interval $[0, 2\pi]$. We construct an RaSEE (we pronounce this ``racy'') state as a weighted sum of the lowest $N_e$ energy eigenvectors.
\begin{equation}
    \Ket{\psi^{\mbox{\tiny RaSEE}}} =
    \sum_k^{N_e} \left| w'_k \right| e^{i \phi_k} \Ket{E_k}
    \label{eq:DefRaSEE}
\end{equation}
\noindent  Figure \ref{fig:ExpRaSEE} shows the values of $S_x$, the expectation value of position, and the  expectation value of the scaled energy for 900 states RaSEE states created using (\ref{eq:DefRaSEE}). We use the same Hamiltonian (with the same connectivity) as that used to calculate the results of Figures \ref{fig:ExpansionEntropies} and \ref{fig:SxMultiNinits}. The figure shows the results of limiting the summation in (\ref{eq:DefRaSEE}) to the lowest $N_e$= 256, 512, or 1024 energy eigenstates, the latter being the full spectrum. The mean value of $S_x$ for all the samples is 9.39 $\pm$ 0.02 bits. Neither the mean value of $S_x$ nor the variance is affected by changing $N_e$. The mean value of $\left< x \right>$ is $1/2$ (centered in the unit square), and as $N_e$ is raised the variance decreases. The mean value of  $\left< E_s \right>$ over the RaSEE samples does increase as larger swaths of the energy spectrum are sampled: $\left< E_s \right>$=13.2 for $N_e$=256, 15.3 for $N_e$=512, and 18.7 for $N_e$=1024.  

\begin{figure}[hbt]
\centering
\includegraphics[width=8.6cm]{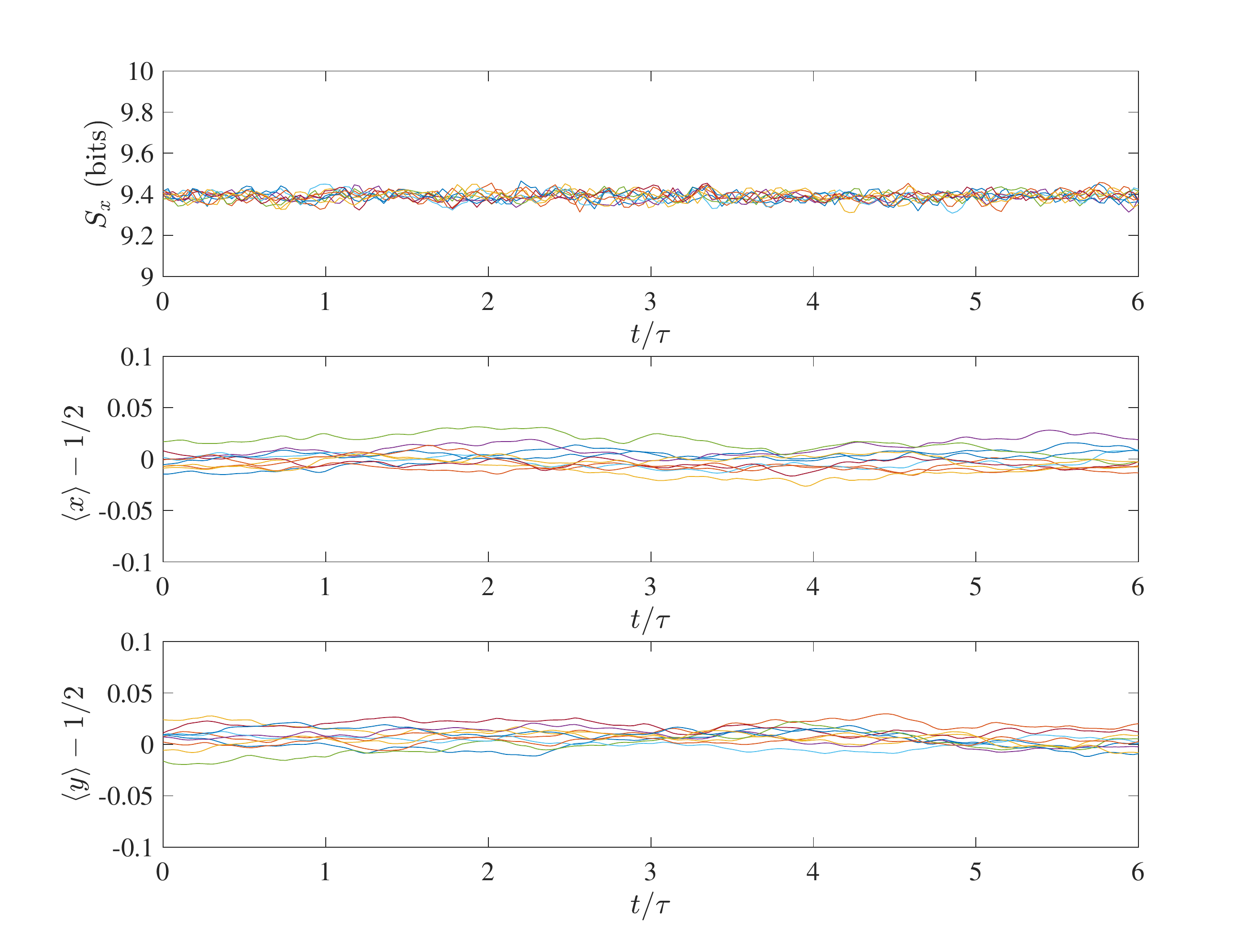}
\caption{Unitary time evolution of RaSEE states. RaSEE states are not stationary states, but exhibit small quantum oscillations around mean values.  This suggests RaSEE states tend to stay in the RaSEE subspace.  Here 10 RaSEE states are used an initial states and evolve under (\ref{eq:UnitaryTimeEvolutionPsi}).  Panel (a) shows the  position entropy $S_x(t)$, and (b) and (c) show the expectation values for $x$ and $y$. 
}
\label{fig:TDSE_RaSEE}
\end{figure}

The RaSEE states are not, of course, stationary states.  Figure \ref{fig:TDSE_RaSEE} shows how $S_x$ and position expectation values vary in time under unitary evolution for ten different RaSEE initial states. Position expectation values $\left< x \right>$  and $\left< y \right>$  fluctuate but stay close to the mean of 1/2. The $S_x$ values also vary in time but remain  close  to the mean value of 9.39 bits. If the system size is doubled to N=2048 (not shown), the mean RaSEE value of $S_x$ becomes 10.39 $\pm$ 0.01 bits-- one more bit of missing position information and an even tighter variance.   

What we see in Figures \ref{fig:SMulticonfig64} and \ref{fig:SxMultiNinits} is that initially localized states, regardless of how localized they are, expand and $S_x(t)$ increases and saturates around a ``typical'' value of 9.39 bits. This value comes from the detailed unitary dynamics for different initial states, but is independent of the details. In fact, we can get a very good estimate of the saturation value  of $S_x$, without having to solve the full dynamics, by simply picking a single RaSEE state using (\ref{eq:DefRaSEE}) and calculating the corresponding $S_x$. If we construct a population of such states we get an even better estimate of the $S_x$ saturation value.  Even if the disorder is varied, so that the energy eigenvalues and eigenstates are slightly different,  as in Figure \ref{fig:SMulticonfig64}, we see the saturation value is  well-approximated by the RaSEE result for any one of the configurations.

The vast majority of RaSEE states have ``typical'' values of $S_x$, the same values to which unitary dynamics drives the system regardless of its initial confinement, as in Figure  \ref{fig:SxMultiNinits}. But RaSEE states are not typical in terms of energy expectation values. For $N_e$=1024 (full spectrum) energy expectation values for RaSEE states are significantly higher than all but the most confined states. 

One might expect that the same  typical value of $S_x$ could be obtained by any unbiased sampling of the surface of the unit sphere in the $N$-dimensional Hilbert space of the the problem, using the eigenstates of any observable. They are connected by a unitary transformation, so all would seem to be equivalent. 
That would be true if we were calculating the expectation value of an observable, which cannot depend on the basis used. But the operator entropy (\ref{eq:SQpure}) is not an expectation value and is in fact a very non-linear function of the state. If we construct, for example a random superposition of {\em position} eigenstates rather than energy eigenstates, we get a typical value of $S_x$= 8.95 bits, which does not match the temporal saturation value. The fact that the saturation value of $S_x(t)$ is accurately generated by sampling the RaSEE states is presumably related to the special role of energy in the time propagation of (\ref{eq:UnitaryTimeEvolutionPsi}).

% 8.95

%If the system simply spread out among the available states with uniformly occupancy, it would relax to $S_x=10$ (because the state space is 1023=$2^{10}$). But it cannot do that without loosing energy, which is impossible under (\ref{eq:UnitaryTimeEvolutionPsi}). 

%%%%%%%%%%% Figure 7  Sx Expectation values of Random Selection  of RaSEE  
%%%%%%%%%%%  %%%%%%%%%%% %%%%%%%%%%% %%%%%%%%%%% 

%%%%%%%%%%% Figure 8  TDSE of RaSEE  
%%%%%%%%%%%  %%%%%%%%%%% %%%%%%%%%%% %%%%%%%%%%% 

%%%%%%%%%%%%%%%%%%%%%%%%%%%%%%%%%%%%%%%%%%%%%%%%%%
%%%%%%%%%%%%%%%%%%%%%%%%%%%%%%%%%%%%%%%%%%%%%%%%%%
\section{Second law of thermodynamics}

%%%%%%%%%%% Figure 9  Blip in time  
\begin{figure}[hbt]
\centering
\includegraphics[width=8.6cm]{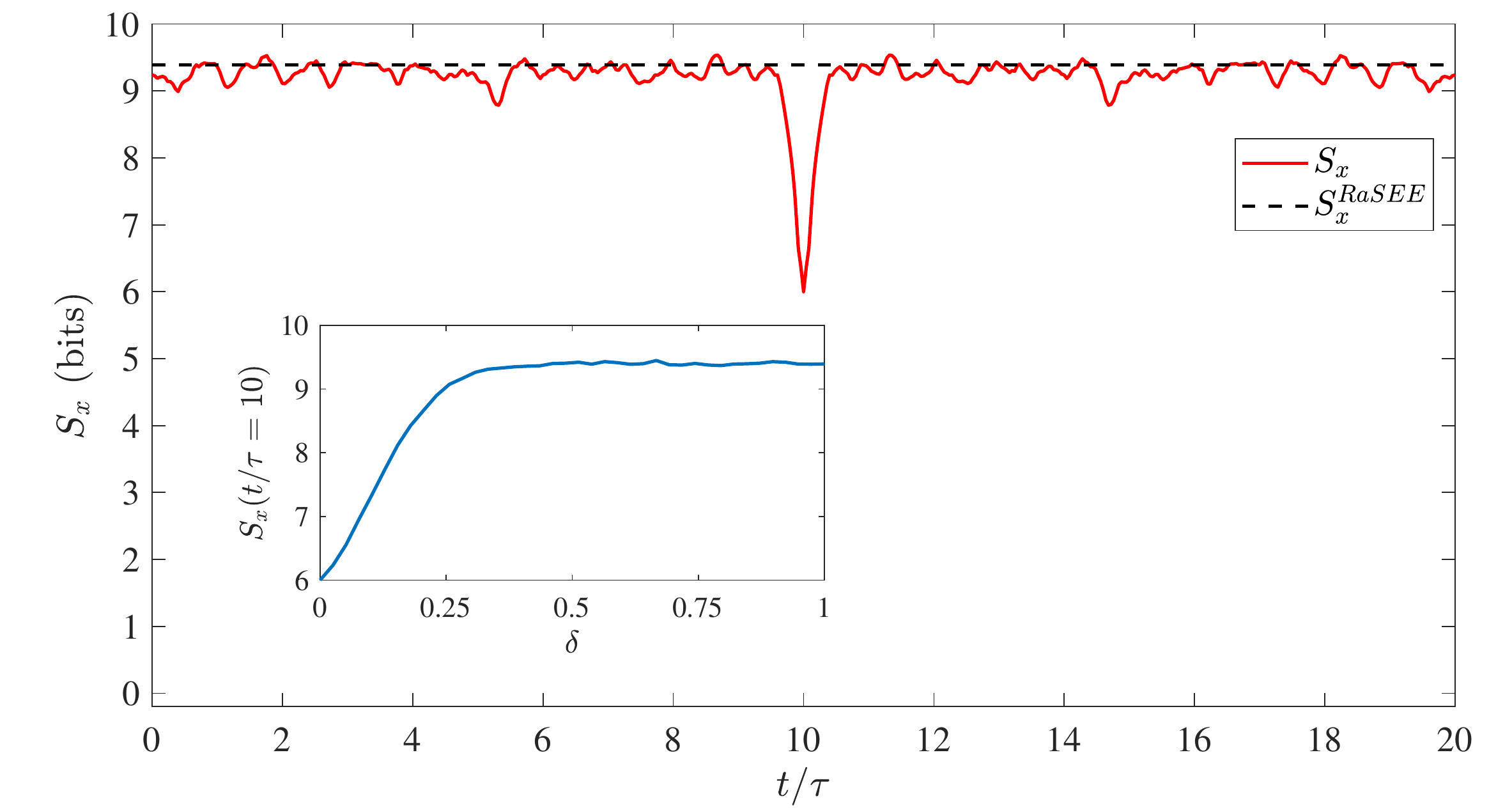}
\caption{Tuning the initial state to achieve momentary localization. We construct a state $\Ket{\psi_0}$ from the time-reversed evolution of the confined state shown in  Figure \ref{fig:ExpansionSnapshots}a and use it as an initial state for subsequent forward time-propagation. Because of the time-symmetry of unitary evolution, there is a spontaneous reduction of the position entropy from the typical $S_x$=9.29 bits to $S_x=6$ bits as the complex interference results in localization at $t=10\tau$. The dashed line shows the position entropy for RaSEE states.  If the initial state is not so carefully prepared, but is 
$\Ket{\psi(0)} \propto \Ket{\psi_0}+\sqrt{\delta}  \Ket{\psi^{\mbox{\tiny RaSEE}}}$,  the entropy reduction is not as great. The inset shows $S_x$ evaluated at the localization time $10\tau$ as a function of $\delta$. As $\delta$ becomes larger, the position entropy  increases to the steady state value and the localization event is destroyed. Small values of $\delta$ do not suppress the localization completely, indicating that the initial state need not be exquisitely tuned to see the momentary violation of the second law of thermodynamics.  
}
\label{fig:BlipInTime}
\end{figure}   
%%%%%%%%%%%  %%%%%%%%%%% %%%%%%%%%%% %%%%%%%%%%% 

The increase in entropy associated with the free expansion of the system we have seen in Figures \ref{fig:ExpansionEntropies} and \ref{fig:SMulticonfig64} suggests a connection to the second law of thermodynamics. We should note that the von Neuman entropy is constant in these cases. Also, many limit entropy as a thermodynamic concept to systems in thermal equilibrium with the environment.  This is not the case here (though see Section \ref{sec:ThermalGroundState})--any of the quantum operator entropies are simply properties of a quantum state determined by (\ref{eq:SQpure}) or (\ref{eq:SQFromRho}) and vary in time as the state varies in time. Each entropy characterizes a quantum state at a particular time (how much information of a particular kind is missing), but an entropy is not an expectation value of an observable. 

As Lesovik has pointed out \cite{Lesovik2013}, the microscopic origin of the increase of entropy can be attributed to the dynamics of the Schr\"odinger equation itself. The spreading of the wave packet, of which the current model is an elaboration, is a feature of the basic structure of the Hamiltonian and the time development operator. 

One then has to deal with the apparent paradox of the time-reversibility of (\ref{eq:UnitaryTimeEvolutionPsi}) in light of the second law. It is helpful to illustrate this  in the current model with its attendant complexity. Let us start with a  confined (localized) state with uniform probability over the 64 sites closes to the origin, such as is shown in Figure \ref{fig:ExpansionSnapshots}a, and which we will denote $\Ket{\psi_{\mbox{\tiny Loc}}}$. Now define a state $\Ket{\psi_0}$ constructed by using the time-reversed version of (\ref{eq:UnitaryTimeEvolutionPsi}), moving backward in time $10 \tau$ with our model Hamiltonian.
\begin{equation}
        \Ket{\psi_0} = e^{+i \hat{H} (10 \tau) /\hbar } \Ket{\psi_{\mbox{\tiny Loc}}}
    \label{eq:BackwardTimeEvolutionPsi}
\end{equation}
\noindent  We now use this state as an initial state, 
$\Ket{\psi(0)}=\Ket{\psi_0}$ and solve (\ref{eq:UnitaryTimeEvolutionPsi}) forward in time. The resultant $S_x(t)$ is shown in Figure \ref{fig:BlipInTime}.
We do indeed see  second-law-violating behavior for $S_x(t)$ inasmuch as there is an abrupt drop in the entropy from its steady state value to a value of 6 bits as the precise combination of magnitudes, phases, and interference combine to reconstruct the localized state at $t=10\tau$. Moreover, the quantum fluctuations prior to and after that point cause $S_x$ to move both up and down around the typical level of 9.39 bits (which is a feature of almost any RaSEE state). So the change in the position entropy moment-by-moment is not monotonically non-negative even away from the sudden recovery of localization.

If one could control all the phases and amplitudes that define the quantum state, nothing in the physical law prevents one from constructing a state like $\Ket{\psi_0}$  that behaves in this second-law-violating way.  Indeed, for a small number of bits Lesovik and coworkers have done just that on the IBM quantum computer \cite{Lesovik2019}. The time-symmetry is broken, not by (\ref{eq:UnitaryTimeEvolutionPsi}) or by a cosmological condition, but by the difficulty of constructing, artificially or naturally, an initial state sufficiently well tuned in phases and magnitudes to produce even the brief and fleeting reduction in entropy we see here. This point was made  by Lesovik in \cite{Lesovik2013} and is underscored by the present model behavior. 

One might suspect that the complex interference that results in the reduction of $S_x$ from about 9.39 to 6 is fragile, in the sense that any small perturbation of the initial state $\Ket{\psi_0}$ would destroy it. To test the resilience of the drop in entropy, we alter the initial state $\Ket{\psi_0}$ by adding to it a fraction of a RaSEE state  and renormalizing.  
\begin{equation}
    \Ket{\psi(0)} = \frac{\Ket{\psi_0}+\sqrt{\delta}  \Ket{\psi^{\mbox{\tiny RaSEE}}} }
     {\left\| \Ket{\psi_0}+\sqrt{\delta}  \Ket{\psi^{\mbox{\tiny RaSEE}}}   \right\|}
\end{equation}
\noindent Here the real scalar $\delta \in [0,1]$ determines the amount of  of the RaSEE state  in the initial state. The inset of Figure \ref{fig:BlipInTime} shows  $S_x(10\tau)$, the  value of the position entropy at the moment of localization recovery, as a function of $\delta$. For small values of $\delta$,  $S_x$ at this minimum increases linearly, recovering to  the steady-state (and RaSEE) value of 9.39 bits by about $\delta=0.3$. Different disordered connectivities produce essentially identical results. Thus, achieving the sudden moment of localization does not in fact require an exquisite tuning of the initial state. Getting the initial state slightly wrong will not wipe out the later localization, just diminish it. 

Nevertheless, it is worth observing that the momentary localization of the state is very brief, lasting only about $\tau$. Thus in the long history of this carefully prepared system, there is a  fleeting blip in time when it spontaneously localizes for a moment. That moment, if achieved, will almost certainly not  be repeated in many lifetimes of the universe.

%%%%%%%%%%%%%%%%%%%%%%%%%%%%%%%%%%%%%%%%%%%%%%%%%%
%%%%%%%%%%%%%%%%%%%%%%%%%%%%%%%%%%%%%%%%%%%%%%%%%%
\section{Thermal ground state \label{sec:ThermalGroundState}}

We have considered the time evolution of a pure quantum state during free expansion. We now briefly describe the application of quantum operator entropies to a state in thermal equilibrium with a reservoir at temperature $T$. In that case, the system degrees of freedom interact with the reservoir degrees of freedom so that energy flows between system and reservoir and their quantum states become entangled. The best local quantum description one can give for the system is then a reduced density matrix (\ref{eq:RhoReduced}) where the unknown reservoir degrees of freedom have been traced out. As Jaynes showed, the optimal reduced density matrix in this case is the one which maximizes the von Neumann entropy over variations of each element of the density matrix \cite{Jaynes1957a}. This is the right entropy to maximize because the effect of the reservoir is precisely to mix system states through entanglement with the reservoir and it is this ``mixedness'' that is quantified by the von Neumann entropy.  The result of the maximum entropy  procedure is the canonical expression for the density operator.  
\begin{equation}
    \hat{\rho} = \frac{e^{-\hat{H}/(k_B T) }} {\Tr \left[ e^{-\hat{H}/(k_B T) }  \right]  }
    \label{eq:ThermalRho}
\end{equation}
The density matix is diagonal in basis of Hamiltonian eigenstates for the system. The von Neumannn entropy is the Shannon entropy of the diagonal elements of $\rho$, and is therefore equal to the  energy  entropy,  $S_E =S_{vN}$.  The off-diagonal elements of the density operator in the position basis do not vanish, and the position entropy $S_x$  remains as a measure of the missing information about position for the state defined by (\ref{eq:ThermalRho}).

\begin{figure}[hbt]
\centering
\includegraphics[width=8.6cm]{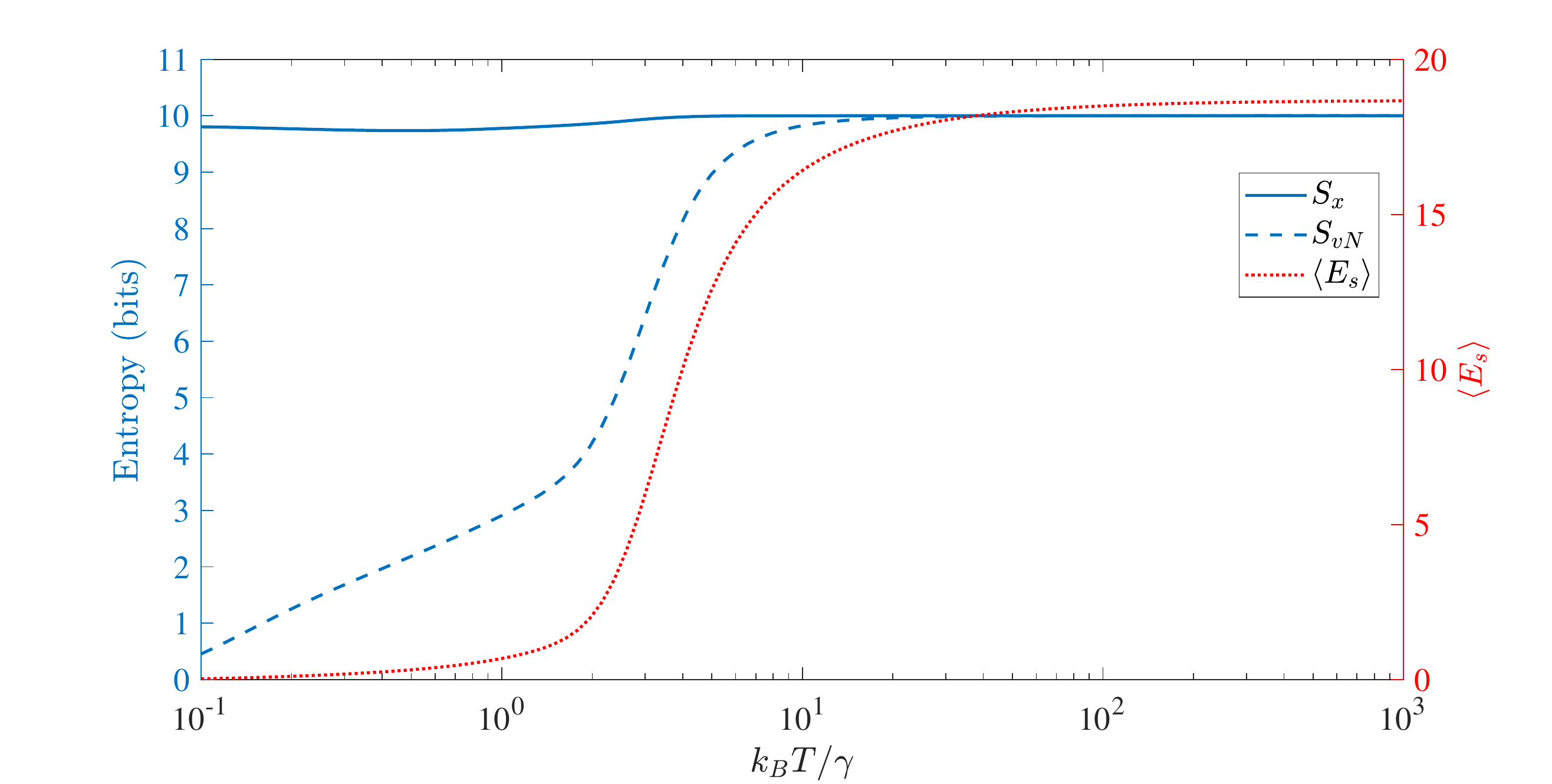}
\caption{Operator entropies and energy expectation values in thermal equilibrium. The position entropy $S_x$, the von Neumann entropy $S_{vN}$, and the scaled energy expectation value $\Braket{E}$ (right axis) are computed from the equilibrium density matrix (\ref{eq:ThermalRho}) at temperature $T$. In the low temperature limit only the ground state is occupied and $S_{vN}\rightarrow{}0$, while $S_x\rightarrow{}9.8$ bits. In the high temperature limit both entropies saturate at 10 bits, consistent with delocalization over all 1024 sites. 
}
\label{fig:ThermalGroundState}
\end{figure}

Figure \ref{fig:ThermalGroundState} shows the position entropy, von Neumann entropy, and the expectation value of the scaled energy $\left< E_s \right>$ as a function of temperature for the model system described by Figures \ref{fig:RandomSiteGeometry}, \ref{fig:ExpansionSnapshots}, and  \ref{fig:ExpansionEntropies}. In the low temperature limit $S_{vN}\xrightarrow{} 0$ and  $\left< E_s \right> \xrightarrow{} 0$ as the system cools to the ground state. The position entropy of the ground state is in this case $S_x=9.8$ bits (notably {\em larger} than the RaSEE value). In the high temperature limit both $S_x$ and $S_{vN}$ go to 10 bits, the completely delocalized state over 1024 sites. 

%%%%%%%%%%% Figure 10  Thermal ground state  
%%%%%%%%%%%  %%%%%%%%%%% %%%%%%%%%%% %%%%%%%%%%% 

% ground state Sx=9.81

%%%%%%%%% Broader interpretation of the model results  %%%%%%%%%%%%%%%
\section{Discussion}

The topologically disordered model described here is of more general applicability than it might at first seem.
It is clear that the actual positions of the sites $\mathbf{r}_k$, the eigenvalues of the position operator, play a limited role in the Hamiltonian (\ref{eq:RandomNetworkH}). These positions yield a simple algorithm for determining the connectivity between the basis states---random choices from among the 50 nearest neighbors---and a way of visualizing the evolution as in Figure \ref{fig:ExpansionSnapshots}. But what really matters is just the connectivity that is shown in Figure \ref{fig:RandomSiteGeometry}b and the coupling strengths.  We could, for example,  re-interpret each basis state $\Ket{\mathbf{r}_k}$ as representing a particular many-body nuclear and electronic configuration for a  molecule. For each configuration there are  a set of other accessible  configurations that are dynamically coupled by the Hamiltonian, giving a sense of ``nearby states,'' but without a regular pattern.   The essential point is that the increase in $S_x(t)$ is capturing quantitatively the very familiar feature of unitary evolution that a localized wavefunction tends to spread out into accessible states, however the states are defined. Nevertheless, in many cases {\em position} eigenstates are especially selected for survival by decoherence through environmental entanglement \cite{Zurek2003b,Blair2013}, so it is not a mistake to focus on them here.  

%But it is helpful to understand that the von Neumann entropy is special case of the more general process of using the Shannon information theoretic measure to determine how much information {\em about a specific observable} is missing from the a  state specificated by $\boldsymbol{\rho}$ (pure or mixed). The quantum operator entropy $S_Q(\boldsymbol{\rho})$ quantifies how many bits of information about $Q$ are missing from $\boldsymbol{\rho}$. The von Neumann entropy is special in that it is the case when the observable $Q$ is the density matrix itself, $S_{vN}(\boldsymbol{\rho})=S_{\rho}(\boldsymbol{\rho})$.

%This is not the same as simply taking the Shannon measure of the diagonal values of the density matrix in an arbitrary representation. Jaynes dismissed proposals by others to define the quantity
%\begin{equation}
%    \sum_k \rho_{k,k} \log_2(\rho_{k,k})
%\end{equation}
%as  the entropy because it depended on the representation chosen. \cite{Jaynes1957b, %TerHaar1955}. This is quite right. Not just any basis set will do.  To be meaningful as a  measure of (missing) information about an actual observable $Q$, the states must be an orthonormal set of eigenstates of $Q$.  Moreover, there are different entropies associated with different quantities and we do not expect that in general the entropy associated with different observables will be equal. 

Different operator entropy measures allows us to capture different aspects of the dynamics. The von Neumann entropy $S_\rho$ captures the ``mixedness'' of a state, which is constant under unitary evolution. We have seen that by contrast the positional entropy $S_x$ increases in a way consistent with the second law of thermodynamics, without recourse to a course-graining procedure. The second law must ultimately be a feature of physical dynamics. There is no law of nature that says systems move from less probable states to more probable states. There is rather just the dynamics of the Schr\"odinger equation applied to the relevant state space.

\end{document}